\documentclass[letterpaper,compsoc,twoside]{IEEEtran}
\usepackage{fixltx2e} 
\usepackage{cmap} 
\usepackage{ifthen}
\usepackage[T1]{fontenc}
\usepackage[utf8]{inputenc}

\usepackage[font={small,it},labelfont=bf]{caption}
\usepackage{float}

\usepackage{longtable,ltcaption,array}
\newlength{\DUtablewidth} 

\pdfoutput=1
\usepackage{scipy}
\makeatletter
\def\PY@reset{\let\PY@it=\relax \let\PY@bf=\relax%
    \let\PY@ul=\relax \let\PY@tc=\relax%
    \let\PY@bc=\relax \let\PY@ff=\relax}
\def\PY@tok#1{\csname PY@tok@#1\endcsname}
\def\PY@toks#1+{\ifx\relax#1\empty\else%
    \PY@tok{#1}\expandafter\PY@toks\fi}
\def\PY@do#1{\PY@bc{\PY@tc{\PY@ul{%
    \PY@it{\PY@bf{\PY@ff{#1}}}}}}}
\def\PY#1#2{\PY@reset\PY@toks#1+\relax+\PY@do{#2}}

\expandafter\def\csname PY@tok@gd\endcsname{\def\PY@tc##1{\textcolor[rgb]{0.63,0.00,0.00}{##1}}}
\expandafter\def\csname PY@tok@gu\endcsname{\let\PY@bf=\textbf\def\PY@tc##1{\textcolor[rgb]{0.50,0.00,0.50}{##1}}}
\expandafter\def\csname PY@tok@gt\endcsname{\def\PY@tc##1{\textcolor[rgb]{0.00,0.27,0.87}{##1}}}
\expandafter\def\csname PY@tok@gs\endcsname{\let\PY@bf=\textbf}
\expandafter\def\csname PY@tok@gr\endcsname{\def\PY@tc##1{\textcolor[rgb]{1.00,0.00,0.00}{##1}}}
\expandafter\def\csname PY@tok@cm\endcsname{\let\PY@it=\textit\def\PY@tc##1{\textcolor[rgb]{0.25,0.50,0.56}{##1}}}
\expandafter\def\csname PY@tok@vg\endcsname{\def\PY@tc##1{\textcolor[rgb]{0.73,0.38,0.84}{##1}}}
\expandafter\def\csname PY@tok@m\endcsname{\def\PY@tc##1{\textcolor[rgb]{0.13,0.50,0.31}{##1}}}
\expandafter\def\csname PY@tok@mh\endcsname{\def\PY@tc##1{\textcolor[rgb]{0.13,0.50,0.31}{##1}}}
\expandafter\def\csname PY@tok@cs\endcsname{\def\PY@tc##1{\textcolor[rgb]{0.25,0.50,0.56}{##1}}\def\PY@bc##1{\setlength{\fboxsep}{0pt}\colorbox[rgb]{1.00,0.94,0.94}{\strut ##1}}}
\expandafter\def\csname PY@tok@ge\endcsname{\let\PY@it=\textit}
\expandafter\def\csname PY@tok@vc\endcsname{\def\PY@tc##1{\textcolor[rgb]{0.73,0.38,0.84}{##1}}}
\expandafter\def\csname PY@tok@il\endcsname{\def\PY@tc##1{\textcolor[rgb]{0.13,0.50,0.31}{##1}}}
\expandafter\def\csname PY@tok@go\endcsname{\def\PY@tc##1{\textcolor[rgb]{0.20,0.20,0.20}{##1}}}
\expandafter\def\csname PY@tok@cp\endcsname{\def\PY@tc##1{\textcolor[rgb]{0.00,0.44,0.13}{##1}}}
\expandafter\def\csname PY@tok@gi\endcsname{\def\PY@tc##1{\textcolor[rgb]{0.00,0.63,0.00}{##1}}}
\expandafter\def\csname PY@tok@gh\endcsname{\let\PY@bf=\textbf\def\PY@tc##1{\textcolor[rgb]{0.00,0.00,0.50}{##1}}}
\expandafter\def\csname PY@tok@ni\endcsname{\let\PY@bf=\textbf\def\PY@tc##1{\textcolor[rgb]{0.84,0.33,0.22}{##1}}}
\expandafter\def\csname PY@tok@nl\endcsname{\let\PY@bf=\textbf\def\PY@tc##1{\textcolor[rgb]{0.00,0.13,0.44}{##1}}}
\expandafter\def\csname PY@tok@nn\endcsname{\let\PY@bf=\textbf\def\PY@tc##1{\textcolor[rgb]{0.05,0.52,0.71}{##1}}}
\expandafter\def\csname PY@tok@no\endcsname{\def\PY@tc##1{\textcolor[rgb]{0.38,0.68,0.84}{##1}}}
\expandafter\def\csname PY@tok@na\endcsname{\def\PY@tc##1{\textcolor[rgb]{0.25,0.44,0.63}{##1}}}
\expandafter\def\csname PY@tok@nb\endcsname{\def\PY@tc##1{\textcolor[rgb]{0.00,0.44,0.13}{##1}}}
\expandafter\def\csname PY@tok@nc\endcsname{\let\PY@bf=\textbf\def\PY@tc##1{\textcolor[rgb]{0.05,0.52,0.71}{##1}}}
\expandafter\def\csname PY@tok@nd\endcsname{\let\PY@bf=\textbf\def\PY@tc##1{\textcolor[rgb]{0.33,0.33,0.33}{##1}}}
\expandafter\def\csname PY@tok@ne\endcsname{\def\PY@tc##1{\textcolor[rgb]{0.00,0.44,0.13}{##1}}}
\expandafter\def\csname PY@tok@nf\endcsname{\def\PY@tc##1{\textcolor[rgb]{0.02,0.16,0.49}{##1}}}
\expandafter\def\csname PY@tok@si\endcsname{\let\PY@it=\textit\def\PY@tc##1{\textcolor[rgb]{0.44,0.63,0.82}{##1}}}
\expandafter\def\csname PY@tok@s2\endcsname{\def\PY@tc##1{\textcolor[rgb]{0.25,0.44,0.63}{##1}}}
\expandafter\def\csname PY@tok@vi\endcsname{\def\PY@tc##1{\textcolor[rgb]{0.73,0.38,0.84}{##1}}}
\expandafter\def\csname PY@tok@nt\endcsname{\let\PY@bf=\textbf\def\PY@tc##1{\textcolor[rgb]{0.02,0.16,0.45}{##1}}}
\expandafter\def\csname PY@tok@nv\endcsname{\def\PY@tc##1{\textcolor[rgb]{0.73,0.38,0.84}{##1}}}
\expandafter\def\csname PY@tok@s1\endcsname{\def\PY@tc##1{\textcolor[rgb]{0.25,0.44,0.63}{##1}}}
\expandafter\def\csname PY@tok@gp\endcsname{\let\PY@bf=\textbf\def\PY@tc##1{\textcolor[rgb]{0.78,0.36,0.04}{##1}}}
\expandafter\def\csname PY@tok@sh\endcsname{\def\PY@tc##1{\textcolor[rgb]{0.25,0.44,0.63}{##1}}}
\expandafter\def\csname PY@tok@ow\endcsname{\let\PY@bf=\textbf\def\PY@tc##1{\textcolor[rgb]{0.00,0.44,0.13}{##1}}}
\expandafter\def\csname PY@tok@sx\endcsname{\def\PY@tc##1{\textcolor[rgb]{0.78,0.36,0.04}{##1}}}
\expandafter\def\csname PY@tok@bp\endcsname{\def\PY@tc##1{\textcolor[rgb]{0.00,0.44,0.13}{##1}}}
\expandafter\def\csname PY@tok@c1\endcsname{\let\PY@it=\textit\def\PY@tc##1{\textcolor[rgb]{0.25,0.50,0.56}{##1}}}
\expandafter\def\csname PY@tok@kc\endcsname{\let\PY@bf=\textbf\def\PY@tc##1{\textcolor[rgb]{0.00,0.44,0.13}{##1}}}
\expandafter\def\csname PY@tok@c\endcsname{\let\PY@it=\textit\def\PY@tc##1{\textcolor[rgb]{0.25,0.50,0.56}{##1}}}
\expandafter\def\csname PY@tok@mf\endcsname{\def\PY@tc##1{\textcolor[rgb]{0.13,0.50,0.31}{##1}}}
\expandafter\def\csname PY@tok@err\endcsname{\def\PY@bc##1{\setlength{\fboxsep}{0pt}\fcolorbox[rgb]{1.00,0.00,0.00}{1,1,1}{\strut ##1}}}
\expandafter\def\csname PY@tok@kd\endcsname{\let\PY@bf=\textbf\def\PY@tc##1{\textcolor[rgb]{0.00,0.44,0.13}{##1}}}
\expandafter\def\csname PY@tok@ss\endcsname{\def\PY@tc##1{\textcolor[rgb]{0.32,0.47,0.09}{##1}}}
\expandafter\def\csname PY@tok@sr\endcsname{\def\PY@tc##1{\textcolor[rgb]{0.14,0.33,0.53}{##1}}}
\expandafter\def\csname PY@tok@mo\endcsname{\def\PY@tc##1{\textcolor[rgb]{0.13,0.50,0.31}{##1}}}
\expandafter\def\csname PY@tok@mi\endcsname{\def\PY@tc##1{\textcolor[rgb]{0.13,0.50,0.31}{##1}}}
\expandafter\def\csname PY@tok@kn\endcsname{\let\PY@bf=\textbf\def\PY@tc##1{\textcolor[rgb]{0.00,0.44,0.13}{##1}}}
\expandafter\def\csname PY@tok@o\endcsname{\def\PY@tc##1{\textcolor[rgb]{0.40,0.40,0.40}{##1}}}
\expandafter\def\csname PY@tok@kr\endcsname{\let\PY@bf=\textbf\def\PY@tc##1{\textcolor[rgb]{0.00,0.44,0.13}{##1}}}
\expandafter\def\csname PY@tok@s\endcsname{\def\PY@tc##1{\textcolor[rgb]{0.25,0.44,0.63}{##1}}}
\expandafter\def\csname PY@tok@kp\endcsname{\def\PY@tc##1{\textcolor[rgb]{0.00,0.44,0.13}{##1}}}
\expandafter\def\csname PY@tok@w\endcsname{\def\PY@tc##1{\textcolor[rgb]{0.73,0.73,0.73}{##1}}}
\expandafter\def\csname PY@tok@kt\endcsname{\def\PY@tc##1{\textcolor[rgb]{0.56,0.13,0.00}{##1}}}
\expandafter\def\csname PY@tok@sc\endcsname{\def\PY@tc##1{\textcolor[rgb]{0.25,0.44,0.63}{##1}}}
\expandafter\def\csname PY@tok@sb\endcsname{\def\PY@tc##1{\textcolor[rgb]{0.25,0.44,0.63}{##1}}}
\expandafter\def\csname PY@tok@k\endcsname{\let\PY@bf=\textbf\def\PY@tc##1{\textcolor[rgb]{0.00,0.44,0.13}{##1}}}
\expandafter\def\csname PY@tok@se\endcsname{\let\PY@bf=\textbf\def\PY@tc##1{\textcolor[rgb]{0.25,0.44,0.63}{##1}}}
\expandafter\def\csname PY@tok@sd\endcsname{\let\PY@it=\textit\def\PY@tc##1{\textcolor[rgb]{0.25,0.44,0.63}{##1}}}

\def\PYZus{\char`\_}
\def\PYZob{\char`\{}
\def\PYZcb{\char`\}}

\def\PYZsh{\char`\#}
\def\PYZpc{\char`\%}

\def\PYZhy{\char`\-}
\def\PYZsq{\char`\'}
\def\PYZdq{\char`\"}

\makeatother

\providecommand*{\DUrole}[2]{%
  \ifcsname DUrole#1\endcsname%
    \csname DUrole#1\endcsname{#2}%
  \else
    \ifcsname docutilsrole#1\endcsname%
      \csname docutilsrole#1\endcsname{#2}%
    \else%
      #2%
    \fi%
  \fi%
}

\ifthenelse{\isundefined{\hypersetup}}{
  \usepackage[colorlinks=true,linkcolor=blue,urlcolor=blue]{hyperref}
  \urlstyle{same} 
}{}

\begin{document}
\newcounter{footnotecounter}\title{Predictive Modelling of Toxicity Resulting from Radiotherapy Treatments of Head and Neck Cancer}\author{Jamie A Dean$^{\setcounter{footnotecounter}{1}\fnsymbol{footnotecounter}\setcounter{footnotecounter}{2}\fnsymbol{footnotecounter}}$%
          \setcounter{footnotecounter}{1}\thanks{\fnsymbol{footnotecounter} %
          Corresponding author: \protect\href{mailto:jamie.dean@icr.ac.uk}{jamie.dean@icr.ac.uk}}\setcounter{footnotecounter}{2}\thanks{\fnsymbol{footnotecounter} The Institute of Cancer Research and The Royal Marsden NHS Foundation Trust, London, UK}, Liam C Welsh$^{\setcounter{footnotecounter}{2}\fnsymbol{footnotecounter}}$, Kevin J Harrington$^{\setcounter{footnotecounter}{2}\fnsymbol{footnotecounter}}$, Christopher M Nutting$^{\setcounter{footnotecounter}{2}\fnsymbol{footnotecounter}}$, Sarah L Gulliford$^{\setcounter{footnotecounter}{2}\fnsymbol{footnotecounter}}$\thanks{%

          \noindent%
          Copyright\,\copyright\,2014 Jamie A Dean et al. This is an open-access article distributed under the terms of the Creative Commons Attribution License, which permits unrestricted use, distribution, and reproduction in any medium, provided the original author and source are credited. http://creativecommons.org/licenses/by/3.0/%
        }}\maketitle
          \renewcommand{\leftmark}{PROC. OF THE 7th EUR. CONF. ON PYTHON IN SCIENCE (EUROSCIPY 2014)}
          \renewcommand{\rightmark}{PREDICTIVE MODELLING OF TOXICITY RESULTING FROM RADIOTHERAPY TREATMENTS OF HEAD AND NECK CANCER}

\setcounter{page}{53}
\newcommand*{\docutilsroleref}{\ref}
\newcommand*{\docutilsrolelabel}{\label}
\AtEndDocument{\cleardoublepage}
\begin{abstract}In radiotherapy for head and neck cancer, the radiation dose delivered to the pharyngeal mucosa (mucosal lining of the throat) is thought to be a major contributing factor to dysphagia (swallowing dysfunction), the most commonly reported severe toxicity. There is a variation in the severity of dysphagia experienced by patients. Understanding the role of the dose distribution in dysphagia would allow improvements in the radiotherapy technique to be explored. The 3D dose distributions delivered to the pharyngeal mucosa of 249 patients treated as part of clinical trials were reconstructed. Pydicom was used to extract DICOM (digital imaging and communications in medicine) data (the standard file formats for medical imaging and radiotherapy data). NumPy and SciPy were used to manipulate the data to generate 3D maps of the dose distribution delivered to the pharyngeal mucosa and calculate metrics describing the dose distribution. Multivariate predictive modelling of severe dysphagia, including descriptions of the dose distribution and relevant clinical factors, was performed using Pandas and SciKit-Learn. Matplotlib and Mayavi were used for 2D and 3D data visualisation. A support vector classification model, with feature selection using randomised logistic regression, to predict radiation-induced severe dysphagia, was trained. When this model was independently validated, the area under the receiver operating characteristic curve was 0.54. The model has poor predictive power and work is ongoing to improve the model through alternative feature engineering and statistical modelling approaches.\end{abstract}

\section{Introduction%
  \label{introduction}%
}

Head and neck cancer is the fifth most common cancer worldwide, with an annual incidence of approximately 500,000 cases globally \cite{Jemal}. Radiotherapy is the primary non-surgical treatment of head and neck cancer and is commonly given in combination with chemotherapy and/or surgery. The aims of the treatment are to achieve loco-regional disease control whilst preserving organ function. Modern radiotherapy techniques allow the radiation dose delivered to the patient to be modulated in order to create highly conformal dose distributions, which minimise the doses delivered to normal tissues in close proximity to the tumour. A typical example of a head and neck radiotherapy treatment plan dose distribution is shown in Figure \DUrole{ref}{treatmentPlan}. However, there are still high rates of toxicity, which reduce patients’ quality of life and limit the amount of dose that can be delivered to the tumour and hence the probability of controlling the disease. The most commonly reported severe radiation-induced toxicity, during and following treatment, is dysphagia (swallowing dysfunction). A variation in the severity of radiation-induced dysphagia between patients is observed. The radiation dose delivered to the pharyngeal mucosa (mucosal lining of the throat) is thought to be a major contributing factor to this side effect \cite{Eisbruch}. Understanding the role of the radiotherapy dose distribution in the onset of dysphagia would allow further improvements to the radiotherapy technique, aiming to reduce the number of patients who experience severe dysphagia.\begin{figure}[]\noindent\makebox[\columnwidth][c]{\includegraphics[width=\columnwidth]{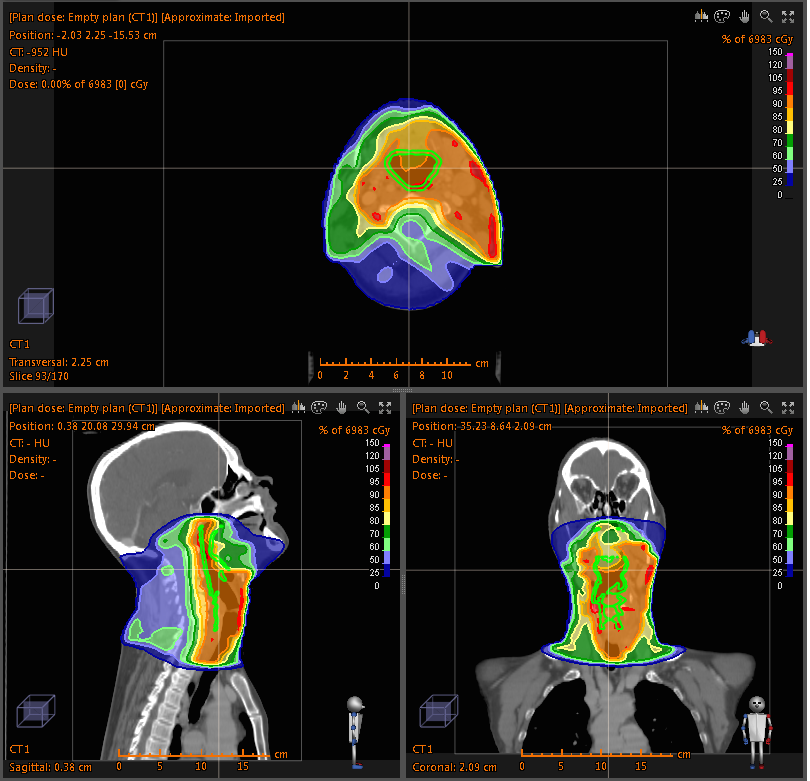}}
\caption{Example of a head and neck treatment plan. The colour wash shows the calculated 3D dose distribution and the pharyngeal mucosa contours, delineated by a radiation oncologist, are shown in green. \DUrole{label}{treatmentPlan}}
\end{figure}

Python allows seamless integration of radiotherapy data handling, manipulation, processing and statistical modelling using machine learning through the Pydicom, NumPy, SciPy, Pandas and SciKit-Learn modules. In this manuscript we outline our application of Python to manipulate radiotherapy treatment plan data to map the planned dose distribution onto the pharyngeal mucosa, compute metrics describing the dose distributions delivered to the mucosa, and train and assess multivariate predictive models of radiation-induced severe dysphagia.

\section{Materials and Methods%
  \label{materials-and-methods}%
}

\subsection{Patient Data%
  \label{patient-data}%
}

Data were available for 249 head and neck cancer patients treated in clinical trials at The Royal Marsden NHS Foundation Trust \cite{Nutting,Guerrero-Urbano,Powell}. All data were anonymised. Toxicity outcome data were collected for patients as part of the trial protocols. The severity of dysphagia was graded, according to the Common Terminology Criteria for Adverse Events \cite{CTCAE}, on a scale from 0 to 5, with higher grades representing higher severity. These toxicity grades were dichotomised into ‘mild’ and ‘severe’ dysphagia, with severe defined as grade 3 or higher. Grade 3 dysphagia corresponds to the patient requiring the insertion of a percutaneous endoscopic gastrostomy (feeding) tube. It requires a surgical procedure and represents a pronounced detriment to a patient’s quality of life and is thus of high clinical importance. This reasoning warranted the decision to dichotomise the outcomes, as well as the choice of threshold. In addition to its clinical importance, the cut point of grade 3 dysphagia is relatively objective compared with other toxicity outcome measures and so minimises noise in the outcome data. Clinical factors that are potentially relevant to dysphagia were also collected as part of the trial protocols. These included patient age, gender, comorbidities and concomitant treatments.

\subsection{Preparation of Radiotherapy Treatment Plan Data%
  \label{preparation-of-radiotherapy-treatment-plan-data}%
}

The pharyngeal mucosa was defined by head and neck radiation oncologists, who contoured the structure on axial slices of computed tomography (CT) scans using either the Pinnacle (Philips Radiation Oncology Systems, Fitchburg, WI), CadPlan (Varian Medical Systems, Palo Alto, CA) or Eclipse (Varian Medical Systems, Palo Alto, CA) proprietary radiotherapy treatment planning software and the structure definition described in \cite{Bhide}. The treatments were planned and the doses calculated by physicists and dosimetrists, following the trial protocols, using the same software.

All scientific computing tasks were performed using Python v.2.7.6 \cite{Python}. A script was used to anonymise the identifying data in the radiotherapy treatment plan files. When treatment planning is performed, the radiation oncologist names the organs that they delineate. This commonly leads to inconsistencies in the names given to the same organ between patients. A script was written to change the name of the pharyngeal mucosa structure from the one assigned by the radiation oncologist (the user selects the name to change, e.g. ‘Pharyngeal Mucosa’, ‘Pharynx’ or ‘pharyngeal mucosa’, from a list of all of the structures) to ‘PM’. This allowed for batch processing of the tasks relating to the radiotherapy treatment plans. The user can pass a list of the patient identification numbers and the name of the organ of interest, in this case ‘PM’, and the dose maps and metrics outlined below are calculated for all patients, without the need for interventions from the user.

\subsection{Generating 3D Dose Maps%
  \label{generating-3d-dose-maps}%
}

Radiotherapy treatment plans are produced using proprietary software. There is a standardised format for radiotherapy treatment plan data, DICOM RT, in which the treatment plan data is exported from the planning software. The information in a treatment plan is saved in four different types of file:%
\begin{itemize}

\item 

\textbf{CT} - These files contain the computed tomography images used to model the patient and calculate the dose distribution in the treatment planning software. There is one file per 2D slice of the 3D image volume acquired.
\item 

\textbf{RTDOSE} - This file contains the calculated dose distribution, described by points on a 3D grid.
\item 

\textbf{RTSTRUCT} - This file contains the coordinates of the contours of the structures delineated by the radiation oncologist.
\item 

\textbf{RTPLAN} - This file contains the parameters of the radiation beams.
\end{itemize}

To study the dose-toxicity response relationship for the pharyngeal mucosa, it was necessary to reconstruct the planned dose to this structure from the DICOM RT files. This was achieved using Pydicom v.0.9.3-1 \cite{Pydicom}, NumPy v.1.8.1-1 \cite{NumPy}, SciPy v.0.14.0-3 \cite{SciPy} and Matplotlib v.1.3.1-9 \cite{Matplotlib}. The resolutions of the dose grids are courser than the image grid upon which the structure contour points are defined. Moreover, the dose grid does not cover the entire image grid, only the volume that is irradiated. Therefore, the dose grid was tri-linearly interpolated to match the resolution of the image grid (tri-linear interpolation does not lead to noticeable distortions of the dose distributions with the acquired dose grid resolution of 2.5 mm x 2.5 mm x 2.5 mm) and the translational offset between the dose grid and image grid calculated using the image coordinates located in the metadata of the RTSTRUCT files. The matplotlib.path.Path.contains\_points() function and the coordinates of the pharyngeal mucosa structure contours were used to determine which voxels were included in the structure. Since the pharyngeal mucosa is a ‘wall’-type structure as opposed to a ‘solid’ volume the numpy.logical\_xor() function was used to exclude those voxels located within one of the contours, but not part of the mucosa structure. This informed a binary mask consisting of ‘1’s for the voxels in the structure and ‘0’s for the voxels not included in the structure. Multiplying the 3D dose grid with this binary mask produced a 3D map of just the dose delivered to the pharyngeal mucosa.

\subsection{Extracting Dose Metrics%
  \label{extracting-dose-metrics}%
}

In most radiotherapy dose-outcome studies, the 3D dose distribution is reduced to a cumulative dose-volume histogram (DVH), describing the volume of an organ that receives each dose level. DVHs were extracted from the 3D dose maps using the NumPy histogram function:\begin{Verbatim}[commandchars=\\\{\},fontsize=\footnotesize]
\PY{k+kn}{import} \PY{n+nn}{numpy} \PY{k+kn}{as} \PY{n+nn}{np}

\PY{k}{def} \PY{n+nf}{dose\PYZus{}volume\PYZus{}histogram}\PY{p}{(}\PY{n}{organDose}\PY{p}{)}\PY{p}{:}
    \PY{l+s+sd}{\PYZdq{}\PYZdq{}\PYZdq{}Computes the cumulative dose\PYZhy{}volume}
\PY{l+s+sd}{        histogram\PYZdq{}\PYZdq{}\PYZdq{}}

    \PY{n}{hist} \PY{o}{=} \PY{n}{np}\PY{o}{.}\PY{n}{histogram}\PY{p}{(}\PY{n}{organDose}\PY{p}{,}
        \PY{n+nb}{range} \PY{o}{=} \PY{p}{(}\PY{l+m+mi}{0}\PY{p}{,} \PY{l+m+mi}{80}\PY{p}{)}\PY{p}{,} \PY{n}{bins} \PY{o}{=} \PY{l+m+mi}{80}\PY{p}{,}
        \PY{n}{density} \PY{o}{=} \PY{n+nb+bp}{True}\PY{p}{)}\PY{p}{[}\PY{l+m+mi}{0}\PY{p}{]}
    \PY{n}{dvh} \PY{o}{=} \PY{l+m+mf}{1.0} \PY{o}{\PYZhy{}} \PY{n}{np}\PY{o}{.}\PY{n}{cumsum}\PY{p}{(}\PY{n}{hist}\PY{p}{)}

    \PY{k}{return} \PY{n}{dvh}
\end{Verbatim}
The mean and maximum doses were also calculated using NumPy.

\subsection{Statistical Modelling%
  \label{statistical-modelling}%
}

The volumes of pharyngeal mucosa receiving cumulative doses in the range 10 Gy (V10) to 80 Gy (V80) in 5 Gy intervals were extracted from the DVHs for inclusion as model covariates. The mean and maximum doses were also used as model inputs. The clinical factors included as covariates were: age, gender, smoking status (smoker at the time of treatment vs. non-smoker at the time of treatment), alcohol status, induction chemotherapy (binary) and concurrent chemotherapy (binary). Data preparation for statistical modelling was carried out using Pandas v.0.14.1-2 \cite{Pandas}.

In order for the model to be able to inform clinical practice it must have high interpretability and low dimensionality. These requirements informed the choice of modelling strategies employed. Feature selection is often challenging when working with high-dimensionality data. An additional complication to the feature selection problem associated with radiotherapy dose-outcome studies is high collinearity between covariates. This is due to the nature of the physical processes that govern how the radiation dose is deposited in the patient. Adjacent dose levels in the DVH are highly correlated making it challenging to distinguish the dose levels that are important for predicting toxicity outcomes from those that merely correlate with the dose levels that cause toxicity. To have the best chance of overcoming this problem a combination of large variation in the DVHs between patients and a suitable feature selection strategy must be employed.

Statistical modelling was performed using SciKit-Learn v.0.15.1-1 \cite{SciKit-Learn}. A subset of 25 patients (10 \% of the entire cohort) were separated out for use as an independent test set of model performance:\begin{Verbatim}[commandchars=\\\{\},fontsize=\footnotesize]
\PY{k+kn}{import} \PY{n+nn}{sklearn}
\PY{k+kn}{from} \PY{n+nn}{sklearn} \PY{k+kn}{import} \PY{n}{cross\PYZus{}validation}

\PY{c}{\PYZsh{} Split data into training set and independent test}
\PY{c}{\PYZsh{}    set}
\PY{n}{xTrain}\PY{p}{,} \PY{n}{xTest}\PY{p}{,} \PY{n}{yTrain}\PY{p}{,} \PY{n}{yTest} \PY{o}{=}
    \PY{n}{cross\PYZus{}validation}\PY{o}{.}\PY{n}{train\PYZus{}test\PYZus{}split}\PY{p}{(}\PY{n}{xData}\PY{p}{,} \PY{n}{yData}\PY{p}{,}
    \PY{n}{test\PYZus{}size} \PY{o}{=} \PY{l+m+mf}{0.1}\PY{p}{,} \PY{n}{random\PYZus{}state} \PY{o}{=} \PY{l+m+mi}{42}\PY{p}{)}
\end{Verbatim}
The model training pipeline consisted of data centring about 0 and scaling to unit variance followed by a feature selection step and, finally, a model fitting step:\begin{Verbatim}[commandchars=\\\{\},fontsize=\footnotesize]
\PY{k+kn}{from} \PY{n+nn}{sklearn.preprocessing} \PY{k+kn}{import} \PY{n}{StandardScaler}
\PY{k+kn}{from} \PY{n+nn}{sklearn.linear\PYZus{}model}
    \PY{k+kn}{import} \PY{n+nn}{RandomizedLogisticRegression}
\PY{k+kn}{from} \PY{n+nn}{sklearn} \PY{k+kn}{import} \PY{n}{svm}

\PY{c}{\PYZsh{} z\PYZhy{}scale data}
\PY{n}{scaler} \PY{o}{=} \PY{n}{StandardScaler}\PY{p}{(}\PY{p}{)}
\PY{c}{\PYZsh{} Feature selection method}
\PY{n}{featureSelection} \PY{o}{=} \PY{n}{RandomizedLogisticRegression}\PY{p}{(}
    \PY{n}{fit\PYZus{}intercept} \PY{o}{=} \PY{n+nb+bp}{False}\PY{p}{)}
\PY{c}{\PYZsh{} Model fitting method}
\PY{n}{modelFitting} \PY{o}{=} \PY{n}{svm}\PY{o}{.}\PY{n}{SVC}\PY{p}{(}\PY{n}{probability} \PY{o}{=} \PY{n+nb+bp}{True}\PY{p}{)}

\PY{c}{\PYZsh{} Create pipeline}
\PY{n}{estimators} \PY{o}{=} \PY{p}{[}\PY{p}{(}\PY{l+s}{\PYZsq{}}\PY{l+s}{scaler}\PY{l+s}{\PYZsq{}}\PY{p}{,} \PY{n}{scaler}\PY{p}{)}\PY{p}{,}
    \PY{p}{(}\PY{l+s}{\PYZsq{}}\PY{l+s}{featureSelection}\PY{l+s}{\PYZsq{}}\PY{p}{,} \PY{n}{featureSelection}\PY{p}{)}\PY{p}{,}
    \PY{p}{(}\PY{l+s}{\PYZsq{}}\PY{l+s}{modelFitting’, modelFitting)]}
\PY{n}{classifier} \PY{o}{=} \PY{n}{Pipeline}\PY{p}{(}\PY{n}{estimators}\PY{p}{)}
\end{Verbatim}
Since the toxicity outcomes were known, supervised learning techniques could be utilised. There is no obvious solution to the problem of collinearity between the variables. Discriminant analysis techniques, such as linear discriminant analysis and quadratic discriminant analysis, cannot robustly handle correlated variables, making them inappropriate. One potential strategy is to remove the collinearity by performing PCA. However, this results in a model with low interpretability and so would not allow the causal features to be determined. An alternative approach is to initially remove features using interpretable dimensionality reduction techniques, for example, univariate feature selection and recursive feature elimination and then remove remaining correlated variables. These approaches, however, may be unstable and suffer from reduced generalisability as different features may be selected with different datasets. Randomised logistic regression (RLR) \cite{Meinshausen} was chosen for feature selection in order to maximise the stability of the selected features. Support vector classification (SVC) with linear and radial basis function kernels \cite{Cortes} was employed for model fitting as this technique is capable of producing high performance, complex (incorporating non-linearity and interactions), yet interpretable, models. A cross-validated grid search, with stratified 5-fold cross-validation, was used over the whole pipeline to tune the hyper-parameters of the models used for feature selection and model fitting on the reduced set of features:\begin{Verbatim}[commandchars=\\\{\},fontsize=\footnotesize]
\PY{k+kn}{from} \PY{n+nn}{sklearn.grid\PYZus{}search} \PY{k+kn}{import} \PY{n}{GridSearchCV}

\PY{n}{featureSelectionParams} \PY{o}{=} \PY{p}{\PYZob{}}
    \PY{l+s}{\PYZsq{}}\PY{l+s}{featureSelection\PYZus{}\PYZus{}C}\PY{l+s}{\PYZsq{}}\PY{p}{:} \PY{p}{[}
    \PY{l+m+mf}{0.01}\PY{p}{,} \PY{l+m+mf}{0.1}\PY{p}{,} \PY{l+m+mf}{1.0}\PY{p}{,} \PY{l+m+mf}{10.0}\PY{p}{,} \PY{l+m+mf}{100.0}\PY{p}{]}\PY{p}{,}
    \PY{l+s}{\PYZsq{}}\PY{l+s}{featureSelection\PYZus{}\PYZus{}scaling}\PY{l+s}{\PYZsq{}}\PY{p}{:} \PY{p}{[}\PY{l+m+mf}{0.25}\PY{p}{,} \PY{l+m+mf}{0.5}\PY{p}{,} \PY{l+m+mf}{0.75}\PY{p}{]}\PY{p}{,}
    \PY{l+s}{\PYZsq{}}\PY{l+s}{featureSelection\PYZus{}\PYZus{}selection\PYZus{}threshold}\PY{l+s}{\PYZsq{}}\PY{p}{:} \PY{p}{[}
    \PY{l+m+mf}{0.3}\PY{p}{,} \PY{l+m+mf}{0.4}\PY{p}{,} \PY{l+m+mf}{0.5}\PY{p}{]}\PY{p}{\PYZcb{}}

\PY{n}{modelFittingParams} \PY{o}{=} \PY{p}{\PYZob{}}
    \PY{l+s}{\PYZsq{}}\PY{l+s}{modelFitting\PYZus{}\PYZus{}C’: [0.01, 0.1, 1.0, 10.0, 100.0],}
    \PY{l+s}{\PYZsq{}}\PY{l+s}{modelFitting\PYZus{}\PYZus{}kernel}\PY{l+s}{\PYZsq{}}\PY{p}{:} \PY{p}{[}\PY{l+s}{\PYZsq{}}\PY{l+s}{linear}\PY{l+s}{\PYZsq{}}\PY{p}{,} \PY{l+s}{\PYZsq{}}\PY{l+s}{rbf}\PY{l+s}{\PYZsq{}}\PY{p}{]}\PY{p}{,}
    \PY{l+s}{\PYZsq{}}\PY{l+s}{modelFitting\PYZus{}\PYZus{}gamma}\PY{l+s}{\PYZsq{}}\PY{p}{:} \PY{p}{[}\PY{l+m+mf}{0.01}\PY{p}{,} \PY{l+m+mf}{0.1}\PY{p}{,} \PY{l+m+mf}{1.0}\PY{p}{]}\PY{p}{\PYZcb{}}

\PY{n}{parameters} \PY{o}{=} \PY{n+nb}{dict}\PY{p}{(}\PY{n}{featureSelectionParams}\PY{o}{.}\PY{n}{items}\PY{p}{(}\PY{p}{)}
    \PY{o}{+} \PY{n}{modelFittingParams}\PY{o}{.}\PY{n}{items}\PY{p}{(}\PY{p}{)}\PY{p}{)}

\PY{n}{skf} \PY{o}{=} \PY{n}{cross\PYZus{}validation}\PY{o}{.}\PY{n}{StratifiedKFold}\PY{p}{(}\PY{n}{yTrain}\PY{p}{,}
    \PY{n}{n\PYZus{}folds} \PY{o}{=} \PY{l+m+mi}{5}\PY{p}{)}
\PY{n}{gridSearch} \PY{o}{=} \PY{n}{GridSearchCV}\PY{p}{(}\PY{n}{clf}\PY{p}{,} \PY{n}{param\PYZus{}grid} \PY{o}{=}
    \PY{n}{parameters}\PY{p}{,} \PY{n}{cv} \PY{o}{=} \PY{l+m+mi}{5}\PY{p}{,} \PY{n}{scoring} \PY{o}{=} \PY{l+s}{\PYZsq{}}\PY{l+s}{roc\PYZus{}auc}\PY{l+s}{\PYZsq{}}\PY{p}{)}
\PY{n}{gridSearch}\PY{o}{.}\PY{n}{fit}\PY{p}{(}\PY{n}{xTrain}\PY{p}{,} \PY{n}{yTrain}\PY{p}{)}
\PY{n}{bestParameters}\PY{p}{,} \PY{n}{bestScore} \PY{o}{=} \PY{n}{gridSearch}\PY{o}{.}\PY{n}{best\PYZus{}params\PYZus{}}\PY{p}{,}
    \PY{n}{gridSearch}\PY{o}{.}\PY{n}{best\PYZus{}score\PYZus{}}

\PY{k}{print} \PY{l+s}{\PYZsq{}}\PY{l+s}{Best Parameters:}\PY{l+s}{\PYZsq{}}\PY{p}{,} \PY{n}{bestParameters}
\PY{k}{print} \PY{l+s}{\PYZsq{}}\PY{l+s}{Best Score:}\PY{l+s}{\PYZsq{}}\PY{p}{,} \PY{n}{bestScore}
\end{Verbatim}
To obtain an unbiased estimate of the model prediction score on new data a nested 5-fold cross-validation was carried out, with area under the receiver operating characteristic (ROC) curve used as the scoring function:\begin{Verbatim}[commandchars=\\\{\},fontsize=\footnotesize]
\PY{c}{\PYZsh{} Nested 5\PYZhy{}fold stratified cross\PYZhy{}validation for model}
\PY{c}{\PYZsh{}    evaluation}
\PY{n}{scores} \PY{o}{=} \PY{n}{cross\PYZus{}validation}\PY{o}{.}\PY{n}{cross\PYZus{}val\PYZus{}score}\PY{p}{(}\PY{n}{gridSearch}\PY{p}{,}
    \PY{n}{xTrain}\PY{p}{,} \PY{n}{yTrain}\PY{p}{,} \PY{n}{cv} \PY{o}{=} \PY{n}{skf}\PY{p}{,} \PY{n}{scoring} \PY{o}{=} \PY{l+s}{\PYZsq{}}\PY{l+s}{roc\PYZus{}auc}\PY{l+s}{\PYZsq{}}\PY{p}{)}
\PY{k}{print} \PY{l+s}{\PYZsq{}}\PY{l+s}{Cross\PYZhy{}validation: ROC AUC =}
    \PY{o}{\PYZpc{}}\PY{l+m+mf}{0.2}\PY{n}{f} \PY{p}{(}\PY{o}{+}\PY{o}{/}\PY{o}{\PYZhy{}} \PY{o}{\PYZpc{}}\PY{l+m+mf}{0.2}\PY{n}{f}\PY{p}{)}\PY{l+s}{\PYZsq{}}\PY{l+s}{ }\PY{l+s}{\PYZpc{}}\PY{l+s}{ (scores.mean(),}
    \PY{n}{scores}\PY{o}{.}\PY{n}{std}\PY{p}{(}\PY{p}{)}\PY{o}{*}\PY{l+m+mi}{2}\PY{p}{)}
\end{Verbatim}
All cross-validations were stratified to reduce variance in the estimate of model performance. Data from the 25 patients in the independent test dataset, not used in any part of the model training process, were used to perform an independent test of model performance:\begin{Verbatim}[commandchars=\\\{\},fontsize=\footnotesize]
\PY{c}{\PYZsh{} Test classifier on independent test data}
\PY{n}{yPred} \PY{o}{=} \PY{n}{gridSearch}\PY{o}{.}\PY{n}{predict}\PY{p}{(}\PY{n}{xTest}\PY{p}{)}
\PY{k}{print} \PY{n}{classification\PYZus{}report}\PY{p}{(}\PY{n}{yTest}\PY{p}{,} \PY{n}{yPred}\PY{p}{)}
\PY{k}{print} \PY{l+s}{\PYZsq{}}\PY{l+s}{ROC AUC: }\PY{l+s}{\PYZsq{}}\PY{p}{,} \PY{n}{roc\PYZus{}auc\PYZus{}score}\PY{p}{(}\PY{n}{yTest}\PY{p}{,} \PY{n}{yPred}\PY{p}{)}
\end{Verbatim}

\section{Results%
  \label{results}%
}
Figure \DUrole{ref}{pharyngealMucosa} shows an example of a 3D map of the dose delivered to the pharyngeal mucosa. Mayavi v.4.3.1-2 \cite{Mayavi} was used for 3D data visualisation. This pharyngeal mucosa dose distribution is typical of head and neck radiotherapy treatments. The region receiving the highest doses is included in the treatment target volume and there is a dose fall-off superiorly and inferiorly of this region.\begin{figure}[]\noindent\makebox[\columnwidth][c]{\includegraphics[width=\columnwidth]{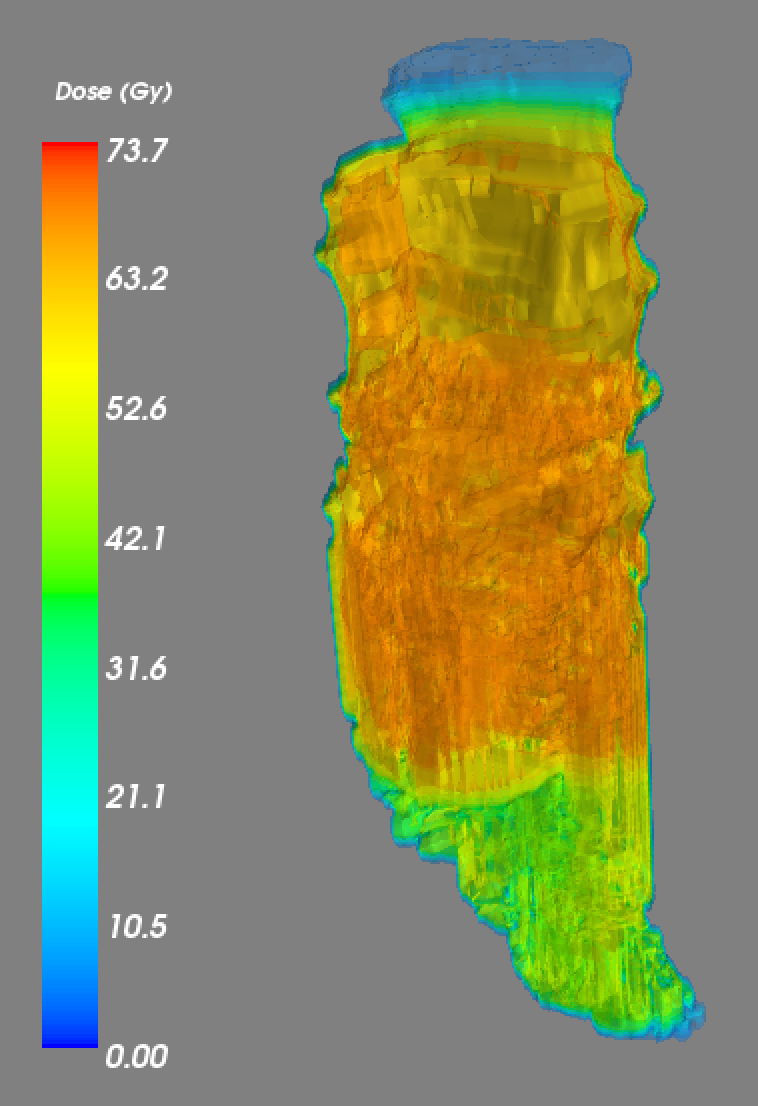}}
\caption{3D map of pharyngeal mucosa dose distribution for one patient. \DUrole{label}{pharyngealMucosa}}
\end{figure}

Figure \DUrole{ref}{dvhs} shows the dose-volume histograms of the 249 patients included in the analysis. Matplotlib was used for 2D data visualisation. It can be observed that there is variation in the pharyngeal mucosa DVHs across the cohort. This variation is largely due to the geometry of the treatment target volume: both its size and location within the pharynx (throat).\begin{figure}[]\noindent\makebox[\columnwidth][c]{\includegraphics[width=\columnwidth]{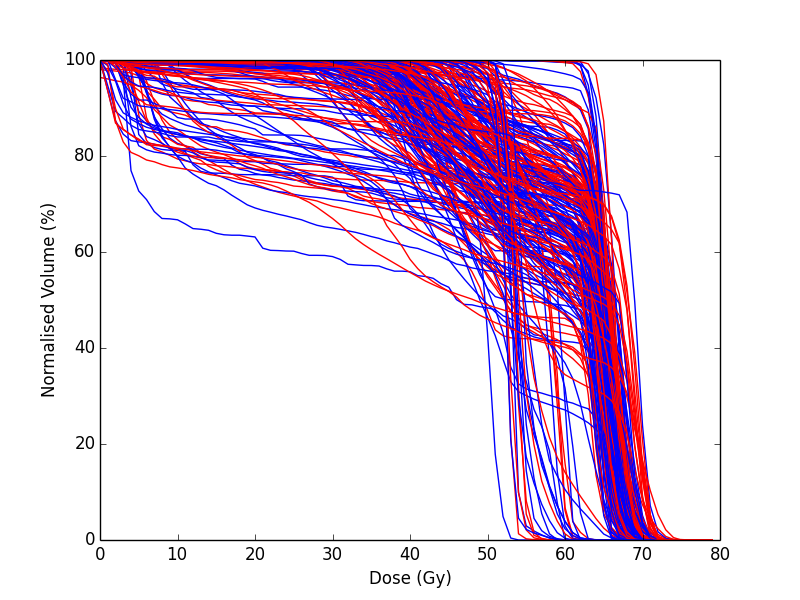}}
\caption{DVHs for the 249 patients analysed. Blue curves represent patients who did not experience severe dysphagia and red curves represent patients who did. \DUrole{label}{dvhs}}
\end{figure}

The correlation matrix for the input variables is shown in Figure \DUrole{ref}{correlationMatrix}. The correlation matrix highlights the high correlation coefficients between adjacent dose levels in the DVH suggesting high collinearity.\begin{figure}[]\noindent\makebox[\columnwidth][c]{\includegraphics[width=\columnwidth]{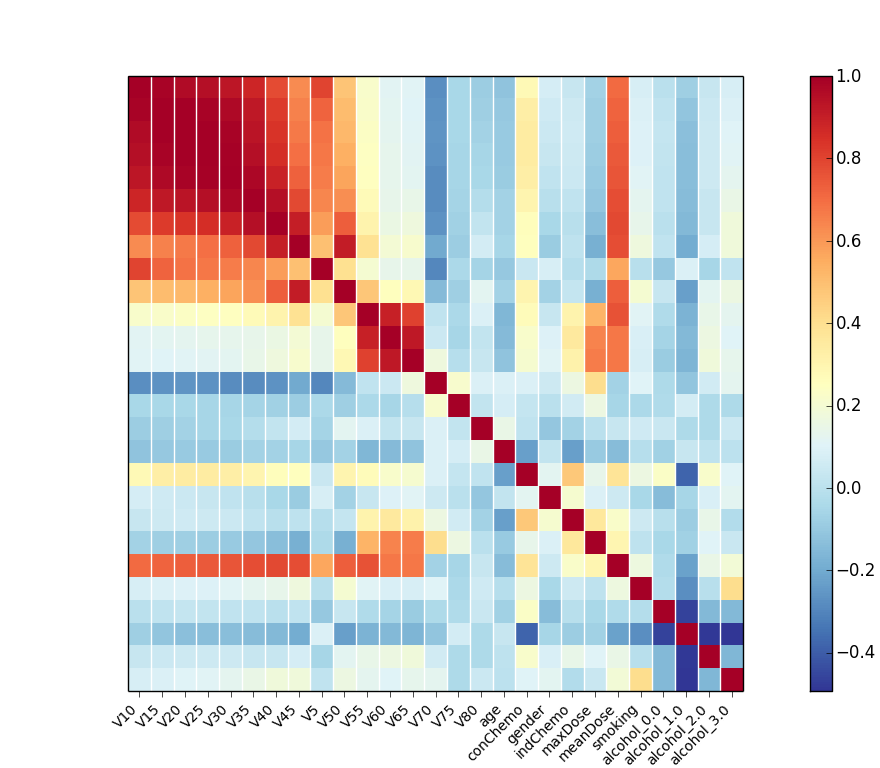}}
\caption{Correlation matrix of model covariates. \DUrole{label}{correlationMatrix}}
\end{figure}

Principal component analysis (PCA) was used for data visualisation. Figure \DUrole{ref}{pcaVarianceExplained} displays the variance explanation and Figure \DUrole{ref}{pca} shows the data projected into the first two principal components space. Visualising the data using PCA shows that the different outcomes are highly overlapping in the first two principal components space (the first two principal components explain a relatively large amount of the variance in the data).\begin{figure}[]\noindent\makebox[\columnwidth][c]{\includegraphics[width=\columnwidth]{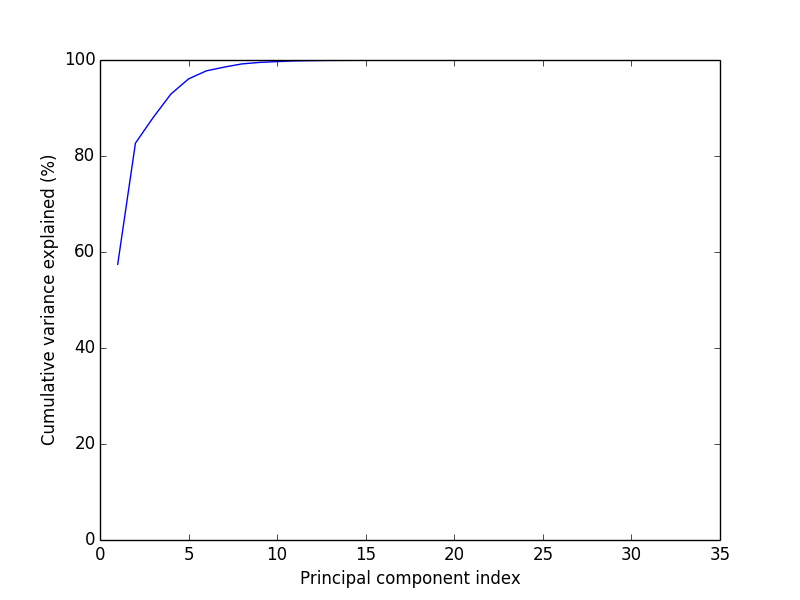}}
\caption{Variance explanation against principal component index from PCA analysis. \DUrole{label}{pcaVarianceExplained}}
\end{figure}\begin{figure}[]\noindent\makebox[\columnwidth][c]{\includegraphics[width=\columnwidth]{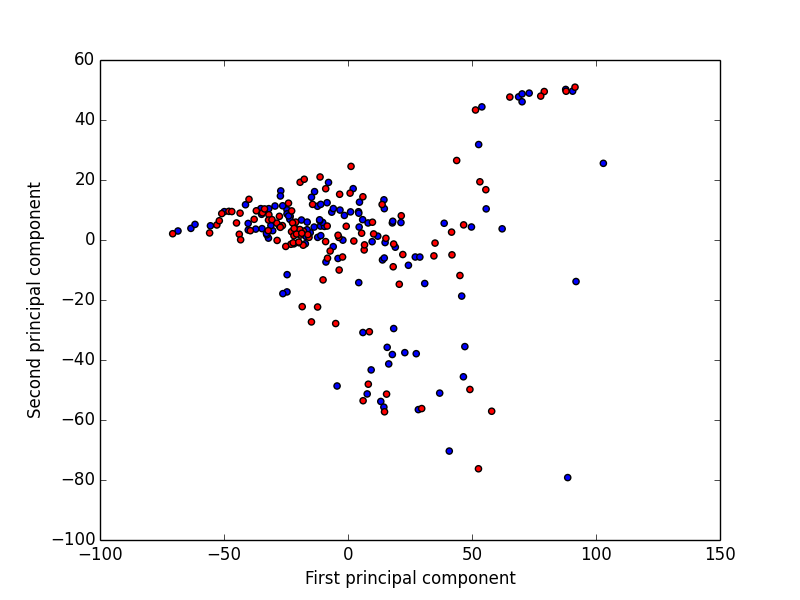}}
\caption{Data projection into the first two principal components space. Red points represent patients who experienced severe dysphagia, blue points represent patients who did not experience severe dysphagia. \DUrole{label}{pca}}
\end{figure}

The model hyper-parameters chosen are shown in Table \DUrole{ref}{hyper-parameters}. During model training the area under the ROC curve was determined to be 0.54 +/- 0.23 and when the model was validated on the independent dataset it was 0.54. Figure \DUrole{ref}{roc} shows the ROC curve for the independent validation.\begin{table}
\setlength{\DUtablewidth}{0.8\linewidth}
\begin{longtable*}[c]{|p{0.435\DUtablewidth}|p{0.284\DUtablewidth}|}
\hline

\textbf{Hyper-parameter} & 

\textbf{Value} \\
\hline

RLR regularisation constant (C) & 

1.0 \\
\hline

RLR threshold & 

0.4 \\
\hline

RLR scaling & 

0.75 \\
\hline

SVC regularisation constant (C) & 

100.0 \\
\hline

SVC kernel & 

Radial basis function \\
\hline

SVC kernel hyper-parameter (gamma) & 

1.0 \\
\hline
\end{longtable*}
\caption{Model hyper-parameters. \DUrole{label}{hyper-parameters}}\end{table}\begin{figure}[]\noindent\makebox[\columnwidth][c]{\includegraphics[width=\columnwidth]{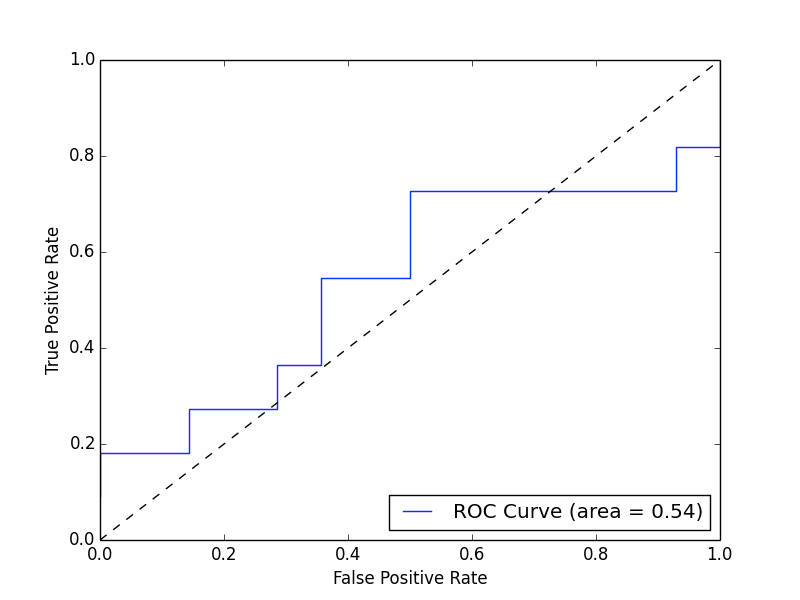}}
\caption{ROC curve for the independent validation. \DUrole{label}{roc}}
\end{figure}

\section{Discussion%
  \label{discussion}%
}

Python was deemed to be a well-suited platform for achieving our aims of training predictive models of severe radiation-induced dysphagia. This is due to its strong functionality in handling DICOM data, manipulation, processing and visualisation of 3D data, and machine learning.

The overlap of the two outcome classes in the first two principal components space upon PCA suggests that generalised linear models, such as logistic regression, are unlikely to make good classifiers in this instance. Non-linear kernel modelling is likely to result in models with greater predictive power for this dataset. During model hyper-parameter tuning, the SVC kernel selected was the radial basis function, supporting this observation.

The area under the ROC curve is low for both the internal and external validation indicating that the model is unable to correctly classify patients better than chance. The uncertainty on the area under the ROC curve from the nested cross-validation is large suggesting that the model is unstable. Our present model is thus unsuitable to support clinical decision making and inform on the causal features of radiation-induced severe dysphagia. However, we are currently exploring improvements to our methodology with promising preliminary results.

There are many potential reasons why our model suffers from low predictive power. These include insufficient characterisation of chemotherapy treatments, smoking status or alcohol status, the influence of other organs, not considered in our study, involved in the onset of severe dysphagia, and the impact of other factors not considered, such as surgery or genetic susceptibility to radiation-induced mucosal damage.

However, we suspect that the low predictive power may, at least in part, be due to limitations associated with the dosimetric input data. In reducing the 3D dose distribution to a DVH much of the information is discarded. We believe that some of this discarded information is likely to be important for predicting toxicity. We are developing several novel dose metrics that more fully characterise the dose distribution, which we aim to use as model inputs in the future. We anticipate that these may improve the predictive power of the models.

Furthermore, there are discrepancies between the planned doses calculated using the treatment planning software (and extracted for use in our study) and the doses that are actually delivered to the patients. These are due to movement of the patients’ internal anatomy whilst the radiation is delivered and weight loss over the course of treatment. In the future we plan to explore the magnitude of these effects and attempt to develop strategies to minimise their impact.

The machine learning pipeline developed has been designed to enable simple addition of different feature selection and model fitting algorithms enabling alternative statistical techniques to be utilised in the future. It also transferable to other toxicities and organs at risk from radiotherapy treatments, for example, lung pneumonitis resulting from radiotherapy for the treatment of lung cancer, as well as studying tumour control.

\section{Conclusions%
  \label{conclusions}%
}

In this study we have shown that Python can be successfully applied to studies of radiotherapy dose-toxicity relationships. The Pydicom, NumPy, SciPy, Pandas and SciKit-Learn modules allow for both manipulation and processing of the treatment planning data, and statistical modelling using machine learning, making Python well suited to this type of study. Whilst initial attempts to generate a predictive model of severe dysphagia were unsuccessful, preliminary investigation of using novel dose metrics to characterise the dose distribution appear promising. Ongoing work involves using NumPy and SciPy to calculate novel dose metrics expected to influence toxicity and the application of alternative statistical methods within the machine learning pipeline.









\begin{thebibliography}{Guerrero-Urbano}
\bibitem[Jemal]{Jemal}{\newcounter{listcnt0}
\begin{list}{\Alph{listcnt0}.}
{
\usecounter{listcnt0}
\setlength{\rightmargin}{\leftmargin}
}

\item 

Jemal et al. \emph{Global cancer statistics}, CA: A Cancer Journal for Clinicians, 61:69-90, 2011.\end{list}
}
\bibitem[Eisbruch]{Eisbruch}{\setcounter{listcnt0}{0}
\begin{list}{\Alph{listcnt0}.}
{
\usecounter{listcnt0}
\setlength{\rightmargin}{\leftmargin}
}

\item 

Eisbruch. \emph{Dysphagia and aspiration following chemo-irradiation of head and neck cancer: major obstacles to intensification of therapy}, Annals of Oncology, 15:363-364, 2004.\end{list}
}
\bibitem[Nutting]{Nutting}{

C.M. Nutting et al. \emph{Parotid-sparing intensity modulated versus conventional radiotherapy in head and neck cancer (PARSPORT): a phase 3 multicentre randomised controlled trial}, Lancet Oncology, 12:127-136, 2011.}
\bibitem[Guerrero-Urbano]{Guerrero-Urbano}{\setcounter{listcnt0}{0}
\begin{list}{\Alph{listcnt0}.}
{
\usecounter{listcnt0}
\addtocounter{listcnt0}{19}
\setlength{\rightmargin}{\leftmargin}
}

\item 

Guerrero Urbano et al. \emph{A phase I study of dose-escalated chemoradiation with accelerated intensity modulated radiotherapy in locally advanced head and neck cancer}, Radiotherapy and Oncology, 85:36-41, 2007.\end{list}
}
\bibitem[Powell]{Powell}{\setcounter{listcnt0}{0}
\begin{list}{\Alph{listcnt0}.}
{
\usecounter{listcnt0}
\addtocounter{listcnt0}{2}
\setlength{\rightmargin}{\leftmargin}
}

\item 

Powell et al. \emph{Fatigue during chemoradiotherapy for nasopharyngeal cancer and its relationship to radiation dose distribution in the brain}, Radiotherapy and Oncology, 110:416-421, 2014.\end{list}
}
\bibitem[CTCAE]{CTCAE}{

The National Cancer Institute. \emph{Common terminology criteria for adverse events v3.0}, 2006.}
\bibitem[Bhide]{Bhide}{

S.A. Bhide et al. \emph{Characteristics of response of oral and pharyngeal mucosa in patients receiving chemo-IMRT for head and neck cancer using hypofractionated accelerated radiotherapy}, Radiotherapy and Oncology, 97:86-91, 2010.}
\bibitem[Python]{Python}{

Python Software Foundation. \emph{Python Language Reference, version 2.7}, Available at \url{http://www.python.org}.}
\bibitem[NumPy]{NumPy}{

T Oliphant et al. \emph{Numerical Python (NumPy)}, \url{http://www.numpy.org} {[}Online; accessed 2014-09-29{]}.}
\bibitem[SciPy]{SciPy}{\setcounter{listcnt0}{0}
\begin{list}{\Alph{listcnt0}.}
{
\usecounter{listcnt0}
\addtocounter{listcnt0}{4}
\setlength{\rightmargin}{\leftmargin}
}

\item 

Jones et al. \emph{SciPy: Open source scientific tools for Python}, \url{http://www.scipy.org/} {[}Online; accessed 2014-08-20{]}, 2001.\end{list}
}
\bibitem[Pydicom]{Pydicom}{\setcounter{listcnt0}{0}
\begin{list}{\Alph{listcnt0}.}
{
\usecounter{listcnt0}
\addtocounter{listcnt0}{3}
\setlength{\rightmargin}{\leftmargin}
}

\item 

Mason. \emph{Pydicom}, \url{https://code.google.com/p/pydicom/} {[}Online; accessed 2014-08-20{]}.\end{list}
}
\bibitem[Pandas]{Pandas}{\setcounter{listcnt0}{0}
\begin{list}{\Alph{listcnt0}.}
{
\usecounter{listcnt0}
\addtocounter{listcnt0}{22}
\setlength{\rightmargin}{\leftmargin}
}

\item 

McKinney. \emph{Data structures for statistical computing in Python}, Proceedings of the 9th Python in Science Conference, 51-56, 2010.\end{list}
}
\bibitem[SciKit-Learn]{SciKit-Learn}{\setcounter{listcnt0}{0}
\begin{list}{\Alph{listcnt0}.}
{
\usecounter{listcnt0}
\addtocounter{listcnt0}{5}
\setlength{\rightmargin}{\leftmargin}
}

\item 

Pedregosa et al. \emph{Scikit-learn: Machine learning in Python}, Journal of Machine Learning Research, 12:2825-2830, 2011.\end{list}
}
\bibitem[Matplotlib]{Matplotlib}{

J.D. Hunter. \emph{Matplotlib: A 2D graphics environment}, Computing in Science and Engineering, 9:90-95, 2007.}
\bibitem[Mayavi]{Mayavi}{\setcounter{listcnt0}{0}
\begin{list}{\Alph{listcnt0}.}
{
\usecounter{listcnt0}
\addtocounter{listcnt0}{15}
\setlength{\rightmargin}{\leftmargin}
}

\item 

Ramachandran and G. Varoquaux. \emph{Mayavi: 3D visualization of scientific data}, IEEE Computing in Science \& Engineering, 13:40-51, 2011.\end{list}
}
\bibitem[Meinshausen]{Meinshausen}{\setcounter{listcnt0}{0}
\begin{list}{\Alph{listcnt0}.}
{
\usecounter{listcnt0}
\addtocounter{listcnt0}{13}
\setlength{\rightmargin}{\leftmargin}
}

\item 

Meinshausen and P. Buhlmann. \emph{Stability selection}, Journal of the Royal Statistical Society: Series B, 72:417-473, 2010.\end{list}
}
\bibitem[Cortes]{Cortes}{\setcounter{listcnt0}{0}
\begin{list}{\Alph{listcnt0}.}
{
\usecounter{listcnt0}
\addtocounter{listcnt0}{2}
\setlength{\rightmargin}{\leftmargin}
}

\item 

Cortes and V. Vapnik. \emph{Support-vector networks}, Machine learning, 20:273-297, 1995.\end{list}
}
\end{thebibliography}
\end{document}